# Crossover Dynamics of Non-Fickian Ionic Diffusion in Solids

Gangbin Yan[1,8,†], Pierfrancesco Ombrini[2,†], Zhichu Tang[3], Shakul Pathak[4], Maoyu Wang[5], Barbara Lavina[5,6], Alexandros Vasileiadis[2], Suin Choi[1], Mingzhan Wang[1], Dongchen Ying[1], Qizhang Li[1], Esen E. Alp[5], Hua Zhou[5], Martin Z. Bazant[4,7], Qian Chen[3], Marnix Wagemaker[2,*] and Chong Liu[1,*]

[1]Pritzker School of Molecular Engineering, University of Chicago, Chicago, IL 60637, USA
[2]Department of Radiation Science and Technology, Delft University of Technology, Delft, 2629JB, the Netherlands
[3]Department of Materials Science and Engineering, the Grainger College of Engineering, University of Illinois Urbana-Champaign, Urbana, IL 61801, USA
[4]Department of Chemical Engineering, Massachusetts Institute of Technology, Cambridge, MA 02139, USA
[5]Advanced Photon Source, Argonne National Laboratory, Lemont, IL 60439, USA
[6]Center for Advanced Radiation Sources, The University of Chicago, Chicago, IL 60637, USA
[7]Department of Mathematics, Massachusetts Institute of Technology, Cambridge, MA 02139, USA
[8]Present address: Department of Physics, Stanford University, Stanford, CA 94305, USA

[†]These authors contributed equally: Gangbin Yan, Pierfrancesco Ombrini
[*]Correspondence: m.wagemaker@tudelft.nl; chongliu@uchicago.edu

**Abstract**

Ionic diffusion in solids is crucial for energy storage[1,2], electronics[3,4], and catalysis[5,6], and thus vital for renewable energy solutions. Yet conventional diffusion models struggle with complexities arising from confinements[7-9], crystallographic disorder[10], lattice distortions[11,12], and coupled transport involving phonons, electrons, or holes[9,13-15]. These challenges are especially pronounced in battery materials, where ionic and electronic carriers move in concert, complicating interpretation of standard electrochemical measurements. Here, we employ tracer exchange as a direct, non-electrochemical probe and reveal rich ion dynamics in the model one-dimensional (1D) conductor olivine $Li_X FePO_4$ ($0 \leq X \leq 1$). $^6Li$-$^7Li$ isotope exchange validates the single-file diffusion (SFD) dynamics, where 1D confinement prevents ion bypassing and preserves spatial order. Kinetic Monte Carlo (KMC) simulations and chronoamperometry measurements further quantify both Faradaic and non-Faradaic surface exchange in nanoplatelets, directly identifying electron transport as the rate-limiting step for electrochemically driven reaction. More strikingly, Li-Na ion exchange exhibits apparent superdiffusion. The exchange rate increases with the increasing Na content. KMC simulations attribute this observation to surface-exchange limitation and enhanced $Li^+$ cross-channel diffusion that drives a dimensional crossover from 1D to quasi-2D transport, which is experimentally supported by four-dimensional scanning transmission electron microscopy (4D-STEM) and in situ synchrotron X-ray diffraction (XRD). Comparisons across particles of different sizes using X-ray absorption spectroscopy (XAS) and Mössbauer spectroscopy further demonstrate that lattice softening facilitates concerted polaron hopping, contributing to the apparent superdiffusive signatures. Our findings offer new insights into ionic diffusion mechanisms in solids, which is made possible by establishing tracer exchange as a powerful tool for probing coupled multi-ion/electron transport.

**Introduction**

Ionic diffusion in solids plays a pivotal role in material science extending from solid-state electrolyte[1,2,10] and mixed ionic-electronic conductors[4,7,16] to perovskites[12] and electrocatalysts[5,6], all critical for next-generation energy systems. The prevailing framework for describing ion transport in solids remains rooted in classical Fickian diffusion, modeled as uncorrelated, memoryless random walks[17]. However, this idealization often fails in materials where transport is shaped by nanoscale confinement, lattice dynamics, or coupling to electronic hopping processes[8,9,17-20] (**Fig. 1a**). For instance, in narrow 1D channels, ions experience self-exclusion, where not every jump attempt is successful, resulting in correlated motion and non-Fickian dynamics[20,21]. In face-centered cubic oxides, concerted multi-ion migration can give rise to superionic conductivity, especially when excess $Li^+$ populates high-energy sites, enhancing $Li^+$-$Li^+$ interactions and altering the energy landscape[14,22]. Furthermore, lattice softening can activate low-energy phonon modes that couple strongly to ion motion, triggering faster migration, correlated hopping or enabling new diffusion pathways[11,23-26]. These examples underscore how atomic-scale structure and dynamics can fundamentally alter solid-state ionic diffusion mechanisms.

Despite growing recognition that diffusion may be anomalous in crystalline solids, the microscopic mechanisms remain poorly understood. A defining feature is the nonlinear scaling of the ensemble-averaged mean-square displacement (eMSD $\propto t^\alpha$, where $\alpha = 1$ corresponds to normal Fickian diffusion; **Fig. 1b**). Scaling analysis reveals distinct transport regimes: superdiffusion ($\alpha > 1$) can arise from occasional long-range displacements, as exemplified by Lévy flights for actively swimming colloids and bacteria[27,28], whereas subdiffusion ($\alpha < 1$) may result from occasional long trapping times or strong spatial confinement, such as in SFD dynamics[18-21,29-37], where particles are constrained from bypassing each other in narrow channels. Although well established in soft matter[38,39] and biological systems[27,28], such frameworks remain largely underexplored in solid-state materials. In fact, continuum models typically assume that tracer diffusivity scales with vacancy concentration but preserve linear eMSD scaling ($\alpha = 1$)[40].

Probing anomalous diffusion of charged species experimentally is especially challenging in materials with coupled ionic and electronic transport. Standard electrochemical techniques conflate these contributions, masking the underlying ion dynamics. To overcome this challenge, here we employ tracer exchange as a direct, non-electrochemical method to isolate ionic transport. Using olivine $Li_XFePO_4$ ($0 \leq X \leq 1$), a model system with strongly anisotropic $Li^+$ diffusion along [010][41-44], we uncover a rich spectrum of diffusive dynamics. Isotopic $^6Li$-$^7Li$ exchange reveals clear SFD characteristics ($\alpha \approx 0.5$), followed by a crossover to normal diffusion ($\alpha \approx 1$) enabled by occasional $Li^+$ passage events through surface or crystal imperfections such as Li-Fe anti-site defects. Importantly, combined KMC simulations and chronoamperometry measurements identify electron transfer as rate-limiting in electrochemically driven processes, directly supporting a coupled ion-electron transfer (CIET) mechanism[13]. More unexpectedly, Li-Na exchange reveals apparent superdiffusion ($\alpha > 1$) in nanoplatelet $Li_XFePO_4$, where surface exchange is the rate-limiting step. Microscopically, $Na^+$-induced in-plane tensile strain reshapes [010] diffusion channels into tapered pathways with distinct $Na^+$ and $Li^+$ preferences at each end. The lattice distortion also promotes quasi-2D $Li^+$ cross-channel diffusion and autocatalytic exchange, whereby the Li-Na exchange accelerates as Na content increases. Moreover, compared with nanoplatelet particles, larger microplatelet particles with more rigid lattices exhibit near-Fickian behavior during Li-Na exchange, further linking superdiffusion to lattice softness and polaron mobility. Together, these findings demonstrate the power of tracer exchange in uncovering complex transport regimes in confined solids and motivate a broader reexamination of ionic diffusion mechanisms beyond classical Fickian models.

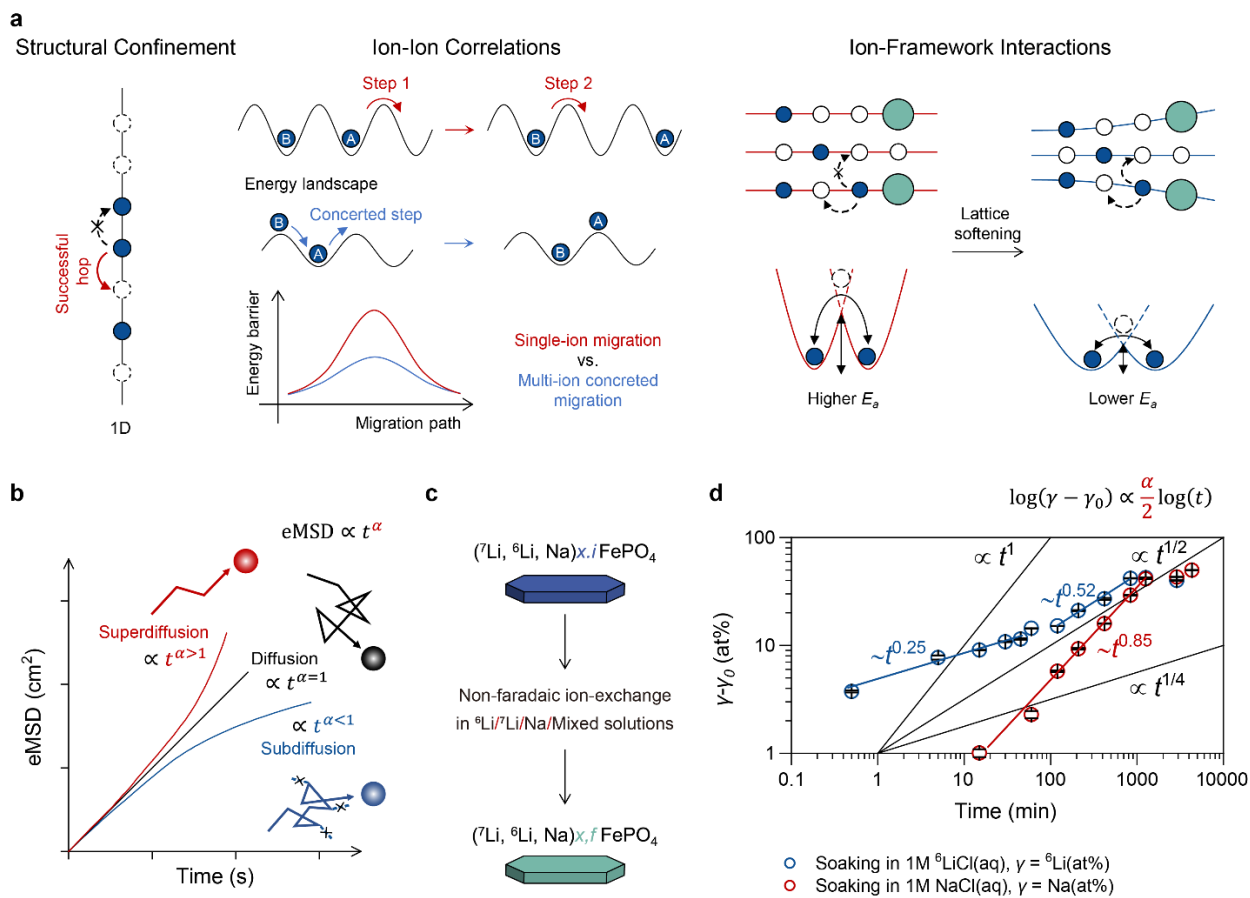

**Fig. 1: Solid-state ionic diffusion and crossover dynamics observed during $^6$Li-$^7$Li self diffusion or Li-Na ion exchange. a**, Structural and chemical factors influencing solid-state ionic diffusion. (Key: lattice frameworks (solid black/red/navy lines); occupied sites by diffusing ions (solid navy/green circles); unoccupied sites (solid/dashed white circles); ion jump attempts (dashed black arrows). **b**, Schematic illustration depicting the eMSD of particle-ensembles over time. For ordinary diffusion or systems follow Fickian diffusion (black), a linear time-dependence ($\alpha = 1$) is expected. Superdiffusive motion (maroon) shows partially or fully directed motion, with a superlinear time-dependence ($\alpha > 1$). In contrast, subdiffusion (blue) indicates slow-than-normal diffusion rates ($\alpha < 1$) due to constrained motions. **c**, Schematic illustrating the experimental design. The compositions of particles (atomic ratios of $^6$Li, $^7$Li, and Na) were measured with Inductively Coupled Plasma Mass Spectrometry (ICP-MS) before and after the non-Faradaic ion exchange in selected tracer ($^6$Li or Na in our cases) solutions. See **Methods** for more experimental details. **d**, Logarithmic plot of tracer exchange atomic ratio change $(\gamma - \gamma_0)$ in 20nm_Li$_{1.0}$FP particles versus time, with blue open circles denote $^6$Li-$^7$Li self diffusion, while red open circles denote Li-Na ion exchange. Specifically, $\gamma_0$ is the initial tracer exchange atomic ratio. The solid blue and red lines are the linear fitting curves, with corresponding fitted time-dependences listed nearby. The black solid lines show the curve following 0.25/0.5/1-time dependence of the tracer exchange ratio. Error bars represent the standard deviation of three replicate measurements.

## Results

*Tracer exchange experiments*

For a typical tracer exchange measurement (**Fig. 1c**), we track the compositions ($^6$Li, $^7$Li, and Na atomic ratios) of the particles over time after soaking them in selected solutions, including 1M $^6$LiCl(aq) and 1M NaCl(aq) (See **Methods** for preparation details). The prepared 1M $^6$LiCl(aq) solution contains 94.8% $^6$Li and 5.2% $^7$Li (**Supplementary Table 1**). High purity $^7$LiOH·H$_2$O with 6.7% $^6$Li is used to synthesize all LiFePO$_4$ particles, and $^6$Li and Na in the external reservoir serve as tracers in our system (**Supplementary Table 1**). The entire process is non-Faradaic, with no electron flow or applied potential.

Here, three distinct well-crystallized LiFePO$_4$ particles with 20 nm, 45 nm, and 340 nm [010] diffusion channel lengths were synthesized using our previous methods[45] (**Supplementary Figs. 1-3**), followed by the surface carbon-coating treatment under calcination (See **Methods** for more synthesis details). Rietveld refinement indicates low anti-site defect levels for all three particles (**Supplementary Figs. 4a, 4d-e** and **Supplementary Table 2**), corroborated by decent capacities (~ 160 mAh/g) delivered during 0.1C (17 mA/g) cycling in 1M LiCl(aq) (**Supplementary Fig. 5**). The $^6$Li, $^7$Li, and Na atomic ratios of all synthesized particles are summarized in **Supplementary Table 3**, with $^6$Li around its natural abundance of 7% and Na content less than 1%. We chose these three types of particles due to their distinct phase evolution behaviors upon electrochemical lithiation, revealed by our in situ synchrotron XRD tracking (**Supplementary Fig. 6**). The 340 nm particles show a first-order phase transition, phase separating between Li-rich and Li-poor phases, while the 20 nm and 45 nm particles exhibit a solid solution transition with a continuous change of lattice parameters, consistent with theoretical predictions of suppressed coherent phase separation in nanoparticles[46,47]. More detailed investigations of phase evolution and morphology quantification are also reported in our previous publication[45]. By spanning the size range for different phase morphologies, this selection ensures a comprehensive analysis of olivine lithium iron phosphate-type particles.

Tracer exchange $(\gamma(t) - \gamma_0)$ is quantified by the fraction of $^6$Li or Na exchanged from the solution over time $t$, where $\gamma_0$ represents the initial atomic ratio of the tracer[48]. Notably, given a 1D diffusion channel length $L$, we can use the relation between the tracer exchange and the eMSD ($(\gamma(t) - \gamma_0)^2 \propto \text{eMSD}(t)/L^2$), to understand the eMSD dynamics in the crystal[18,33,34]. Therefore, by extracting the slope of $\log(\gamma - \gamma_0)$ versus $\log(t)$, which is equal to $\alpha/2$, we can identify different diffusion mechanisms (**Supplementary Note 1**).

With the tracer exchange experiment, we discovered two types of anomalous diffusion. For example, as shown in **Fig. 1d**, 20nm_Li$_{1.0}$FP particles initially exhibit subdiffusion ($\alpha/2 \approx 0.25$) during $^6$Li-$^7$Li self diffusion, which then transition to ordinary diffusion ($\alpha/2 \approx 0.52$). Meanwhile, apparent superdiffusion ($\alpha/2 \approx 0.85$) is observed during Li-Na ion exchange. In the following two sections, we delve into the underlying mechanisms behind the observed non-Fickian ion exchange behavior in the 1D channels of the structures, supported by both experimental and computational evidence.

*Subdiffusion observed in non-Faradaic $^6$Li-$^7$Li self diffusion*

As shown in **Fig. 2a-b** and **Extended Fig. 1a-b**, ubiquitous subdiffusion behavior with a ~0.25-time dependence of $^6$Li exchange ratio is observed for all particles investigated: two batches of carbon-coated 20nm_Li$_{1.0}$FP nanoplatelets, bare uncoated 20nm_Li$_{1.0}$FP_bare nanoplatelets, carbon-coated 45nm_Li$_{1.0}$FP nanoplatelets, and carbon-coated 340nm_Li$_{0.24}$FP microplatelets (**Supplementary Figs. 4 and 7** and **Supplementary Tables 2-3**; See **Methods** for synthesis details).

The observed subdiffusive ($\alpha \approx 0.5$) time-dependence of $^6$Li exchange ratio is closely linked to the intrinsic SFD dynamics of olivine LFP structures. For ordinary 1D Brownian Diffusion (1D-BD), while macroscopic displacement occurs only along the channel axis, the microscopic motions of particles remain 3D (**Extended Fig. 1c**), allowing particle passages. The time dependence of the mean displacement in these systems follows the well-known Einstein equation[17]:

$$\text{eMSD}_{1D-BD} = 2Dt \qquad (1)$$

where $D$ is the self diffusion constant ($[D] = \text{m}^2\text{s}^{-1}$). In contrast, in SFD the diffusion channel only allows single-file passage of Li ions, preventing ion passage events and preserving the ion sequence. Consequently, migration of an ion requires vacancy migration to it. When this vacancy migration follows a standard diffusive process, this results in a subdiffusive ($t^{1/2}$) scaling of the mean displacement[21,29-31,49-53]:

$$\text{eMSD}_{SFD} = 2Ft^{1/2} \qquad (2)$$

where the mobility factor $F(t)$ ($[F] = \text{m}^2\text{s}^{-1/2}$) is introduced. This theoretical prediction has been experimentally validated through microrheology studies in zeolites[31,32,35], confined colloidal systems[34], and water uptake in carbon nanotubes[54,55]. Recent advancements have extended these frameworks by incorporating thermodynamic refinements, including activity coefficients that vary with concentration profiles and account for excluded-site effects in olivine Li$_X$FePO$_4$[40,56]. However, these models still fail to capture the distinct time evolution characteristic of SFD.

More complicated features are witnessed at long times during $^6$Li-$^7$Li self diffusion. The tracer exchange ratio exhibits a transition from 0.25- to 0.5-time dependence, ultimately recovering Fickian behavior. This crossover is expected in finite systems, where sustained anomalous diffusion is not possible[57]. In our case, the transition can arise from occasional particle-passing events, which disrupt particle ordering and break the single-file constraint[33]. As a result, dynamics shift from genuine SFD ($\alpha/2 \approx 0.25$) toward ordinary diffusion ($\alpha/2 \approx 0.5$). The sustained initial 0.25 slope in our data indicates a low passage rate and minimal impact on $\gamma$ at early stages. At longer times, collective motion of the isotopic ensemble dominates the exchange and follows an ordinary diffusion behavior, as predicted by the theory[33].

To quantitatively and microscopically understand the observed $^6$Li-$^7$Li subdiffusion behavior, we used KMC simulations with realistic constraints to explore the effects of various processes involved, including the surface exchange rate between structure and solution ($\Gamma_S$), the bulk diffusion rate of Li hopping to an adjacent site ($\Gamma_B$), and the cross-channel diffusion rate in the correspondence of Li-Fe anti-site defects ($\Gamma_C$) (**Fig. 2c**; see **Supplementary Note 2** and **Supplementary Figs. 8-10** for the KMC simulation algorithm and parameters optimization details). Simulations were focused on 20nm_Li$_{1.0}$FP exchange to avoid local intra-channel or mosaic phase separation.

We captured the SFD dynamics and time dependence transition for 20nm_Li$_{1.0}$FP particles with and without carbon coating, as shown in **Fig. 2a-b**. For both cases, $\Gamma_B$ and $\Gamma_C$ are fitted to be $10^5 \text{ s}^{-1}$ and $1 \text{ s}^{-1}$, respectively. These values correspond to a bulk diffusion energy barrier of ~0.4 eV along [010] channels and a cross-channel activation barrier of 0.71 eV (See **Supplementary Note 3** for calculation details), aligning well with DFT-computed Li-vacancy self-diffusion barriers along [010] channels (0.39 ~ 0.45 eV)[44] and a cross-channel vacancy diffusion activation barrier of 0.7 ~ 0.8 eV under a concerted Li ions diffusion manner[58]. The KMC model further provides insight into key features of the experimental results (**Figs. 2d-e** and **Supplementary Videos 1-2**). Initially, $^6$Li-$^7$Li exchange is dominated by surface interactions between the particle and the reservoir. After ~5 mins, surface concentrations equilibrate with the reservoir, as also highlighted by the strong sensitivity of $\Gamma_S$ in the early time regime (**Supplementary Fig. 8**). The kinetics then enter a SFD regime, where defect-free channels exchange Li isotopes through double open boundaries. This one-dimensional, volume-excluded diffusion follows a characteristic 0.25-time dependence of $^6$Li exchange ratio. Finally, once the defect-free channels are exchanged (~200 minutes; **Fig. 2d**), particle-passing events start to dominate the process, shifting it toward 2D Fickian diffusion with a 0.5-time dependence.

Therefore, the KMC simulations, reproducing the transition from 0.25 to 0.5, comprehend both surface exchange ($\Gamma_S$) and cross-channel diffusion ($\Gamma_C$), which can enable such events (Illustrated in **Extended Fig. 1d**). However, our simulations indicate that defect-assisted cross-channel hopping is the primary factor, as highlighted by the influence of $\Gamma_C$ on the 0.25 to 0.5 scaling transition (**Supplementary Fig. 9**). It is also reported that, the anti-site defects reduce ion mobility along the channels while enhancing intra-channel diffusivity[44]. We can now conclude that these defects not only alter the effective ionic diffusivity of the system, but also the diffusion dynamics itself. An increased defect concentration drives the system away from SFD towards Brownian motion.

Additionally, the strong sensitivity to the surface exchange frequency allows us to accurately quantify its non-Faradaic rate $\Gamma_S$ (**Supplementary Fig. 8**) and decouple it from the one obtained with electrochemical experiments. Despite many studies show that carbon-coated LFP have higher reaction rate ($k_0$) under Faradaic conditions[13,40,59], our non-Faradaic tracer-exchange measurements show that 20nm_Li$_{1.0}$FP_bare particles exhibit faster isotopic exchange (higher $\Gamma_S$) than carbon-coated particles. Specifically, $\Gamma_S$ for the bare 20nm_Li$_{1.0}$FP particles is one order of magnitude higher, likely because the carbon overlayer introduces an additional interfacial diffusion barrier. This inference is supported by inverse learning of LFP reaction kinetics from nanoparticle X-ray imaging experiments, which show that lower intercalation rates are correlated with thicker regions of carbon coating[59]. However, this apparent discrepancy does not contradict with the recently proposed coupled ion-electron transfer (CIET) framework[60], since the governing driving forces differ between conditions with net redox (Faradaic) and those without (non-Faradaic). To investigate the difference between Faradaic and non-Faradaic phenomena (i.e. ion transfer and CIET) we performed chronoamperometry experiments on 20nm_Li$_{1.0}$FP particles w/wo carbon coating. The Faradaic exchange frequencies were quantified using Marcus-Hush-Chidsey (MHC) model, which is equivalent to the Electron-Coupled Ion Transfer (ECIT) limit of CIET theory (aside from temperature dependence)[13], yielding values of $k_{0,bare} = 0.12 \text{ s}^{-1}$ and $k_{0,coated} = 0.49 \text{ s}^{-1}$ for bare and carbon-coated particles, respectively (**Fig. 2f**; See **Supplementary Note 4** for calculations). The fourfold higher rate observed for the carbon-coated particles suggests that bare particles only utilize a fraction of their reactive surface area — specifically, regions in direct contact with conductive carbon within the microstructure. Additionally, the CIET theoretical framework allows us to assess the relative contributions of Faradaic and non-Faradaic mechanisms to the exchange frequency (**Supplementary Note 5**). Based on the theory, we find the following relation:

$$k_0 = \frac{\Delta_e}{h} \frac{\Gamma_S}{v_s} f(c, \lambda) \qquad (3)$$

where $v_s$ is the attempt frequency of the ions, $h$ is the Planck constant, $\Delta_e/h$ is relating to the quantum tunneling rate of electrons[13] and $f(c, \lambda)$ is a function of the solid filling fraction $c$ and reorganization energy $\lambda$ in the Faradaic process. Inserting $\Gamma_S = 10^2 \text{ s}^{-1}$ from the carbon-coated particles, we can reach $k_0 \sim 0.48 \text{ s}^{-1}$, by considering $\Delta_e \sim 10^{-3}$ eV for weakly adiabatic transition, $v_s \sim 10^{12} \text{ s}^{-1}$ in the range of phonon frequency[44], $\lambda = 213$ meV[60], and an average filling fraction of 0.5. The strong agreement between the predicted and the measured exchange frequency further validates the theoretical framework, also confirming that the electrochemically driven Faradaic reaction in the bare particles occurs predominantly at the carbon-coated interfaces or through the surface areas in direct contact with the carbon black.

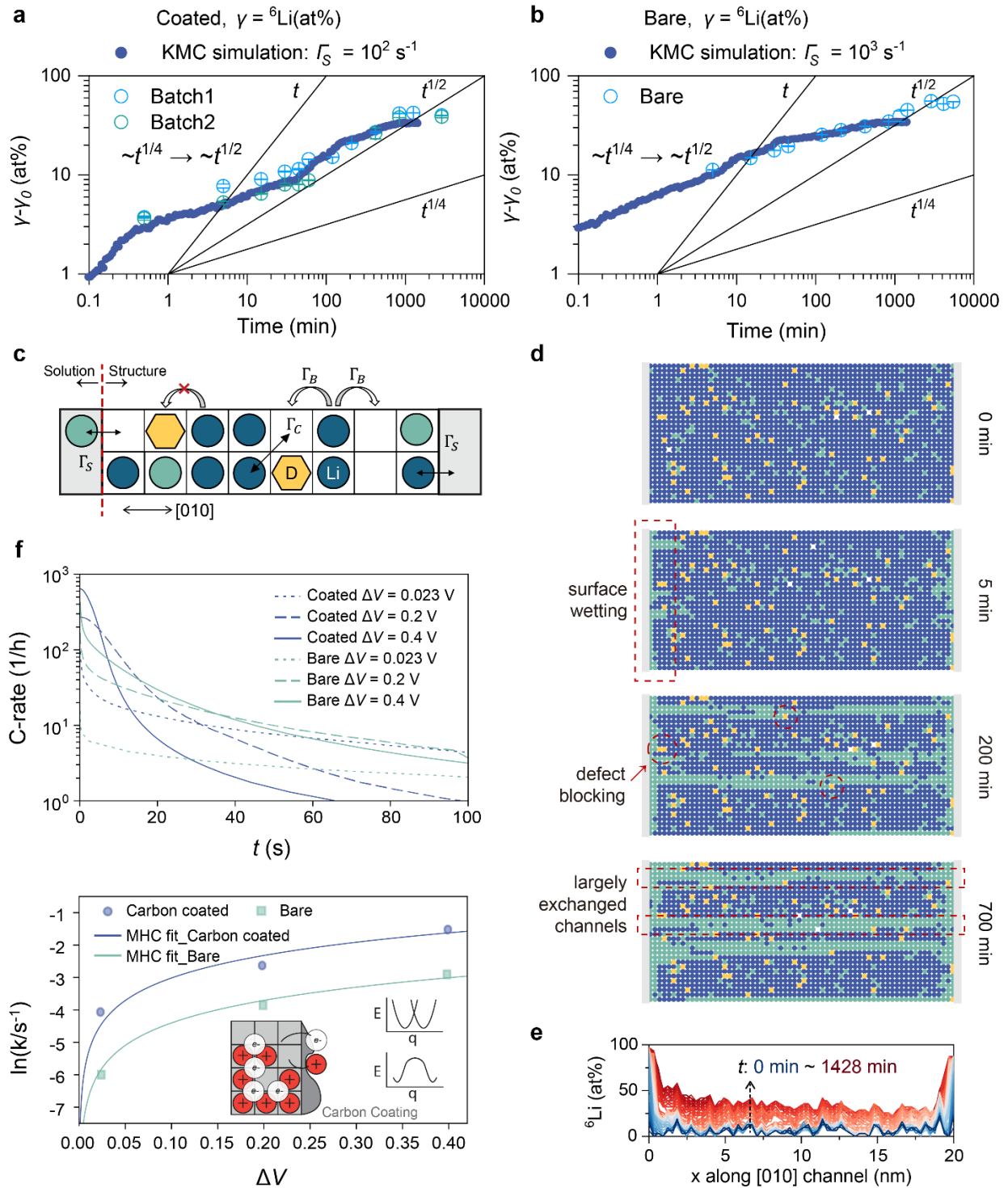

**Fig. 2: Non-Faradaic $^6$Li-$^7$Li self diffusion. a,** Logarithmic plot of $^6$Li tracer atomic ratio changes $(\gamma - \gamma_0)$ over time during $^6$Li-$^7$Li self diffusion in two different batches of carbon-coated 20nm_Li$_{1.0}$FP particles: blue open circles for Batch 1, green open circles for Batch 2. The smaller solid blue dots represent $^6$Li tracer exchange obtained from KMC simulations (Fitted parameters: $\Gamma_B = 10^5$ s$^{-1}$; $\Gamma_S = 10^2$ s$^{-1}$; $\Gamma_C = 1$ s$^{-1}$). **b,** Logarithmic plot of $^6$Li tracer atomic ratio changes $(\gamma - \gamma_0)$ over time during $^6$Li-$^7$Li self

diffusion in bare 20nm_Li$_{1.0}$FP particles without carbon coating (Blue open circles; See **Methods** for more synthesis details). The smaller solid blue dots represent $^6$Li tracer exchange obtained from KMC simulations (Fitted parameters: $\Gamma_B = 10^5 \text{ s}^{-1}$; $\Gamma_S = 10^3 \text{ s}^{-1}$; $\Gamma_C = 1 \text{ s}^{-1}$). **c**, Illustration of events in the KMC simulation: Each square white box represents a Li site in the channel, with original structure Li ions shown as dark blue spheres. Space outside the boxes represents the electrolyte. $\Gamma_S$, $\Gamma_B$, and $\Gamma_C$ correspond to the surface exchange rate between structure Li$^+$ (dark blue spheres) and solution Li$^+$ (light green spheres), bulk diffusion rate of Li$^+$ hopping to an adjacent site, and cross-channel diffusion rate due to Li-Fe anti-site detects (yellow hexagons), respectively. See **Supplementary Note 2** for more KMC simulation details. **d**, Snapshots of KMC simulations for carbon-coated 20nm_Li$_{1.0}$FP particles during $^6$Li-$^7$Li ion exchange. Initially (at $t = 0$ min), ~7% of the storage sites contain naturally abundant $^6$Li. Key: $^7$Li ions (dark blue); $^6$Li ions (light green); defects (yellow). See **Supplementary Video 1** for the complete simulation. **e**, Evolution of $^6$Li tracer atomic ratio for carbon-coated 20nm_Li$_{1.0}$FP particles during $^6$Li-$^7$Li ion exchange calculated from KMC simulations, plotted along the [010] diffusion channel. **f**, Time evolution of normalized current (C-rate) for bare and carbon-coated 20nm_Li$_{1.0}$FP particles under various overpotentials, as measured by chronoamperometry (See **Methods** for more details). The corresponding Tafel plots and MHC model fits are shown below (**Supplementary Note 4**). The schematic illustrates the transition state energy landscape for electron and ion transfer. Error bars in (**a-b**) represent the standard deviation of three replicate measurements.

*Superdiffusion observed in non-Faradaic Li-Na ion exchange*

Apparent superdiffusive behavior was observed during Li-Na ion exchange in 20nm_Li$_X$FP ($X$ = 0.28, 0.48, 0.76, and 1.0) and 45nm_Li$_{1.0}$FP nanoplatelets (**Figs. 3a-b**, and **Extended Figs. 2a-b**). The exact compositions of chemically prepared 20nm_Li$_X$FP ($X$ = 0.28, 0.48, and 0.76) were verified through XRD refinement and ICP-MS (**Supplementary Fig. 11**, **Supplementary Tables 4 and 5**; See **Methods** for more synthesis details). Notably, their lattice parameters and unit cell volumes follow a linear combination of the two end phases, 20nm_FP and 20nm_Li$_{1.0}$FP, again confirming the solid solution phase evolutions upon lithiation (**Supplementary Figs. 6 and 12**) and allowing us to interpret the data considering homogeneous lithiation.

To quantify the microscopic processes underlying Li-Na exchange, we performed KMC simulations on carbon-coated 20nm_Li$_{1.0}$FP particles. Using the previously defined model, we were unable to reproduce the experimentally observed ~0.85 time dependence of Na exchange ratio (**Fig. 3a**). In particular, we found that altering $\Gamma_B$ had a negligible effect on the simulation results (**Supplementary Fig. 10**), since bulk diffusion is already orders of magnitude greater than $\Gamma_S$ and is not rate-limiting. Meanwhile, although $\Gamma_S$ is the dominant rate-limiting factor and strongly influences the overall kinetics of the system, modifying it alone failed to achieve a precise fit (**Supplementary Fig. 10**).

We therefore extended the model by considering the role of Na$^+$ in lattice expansion. We included in the simulation an Na$^+$-assisted Li$^+$ inter-channel diffusion ($\Gamma_{I\_ass}$) along [001], but not along [100], which is blocked by PO$_4$ tetrahedra[61,62]. While microscopically the intra-channel hopping of ions in a distorted lattice might appear quasi-3D, our goal was to evaluate the effect of an additional degree of freedom. To account for the collective long-range behavior of Na$^+$ (further discussed later), we assumed the rate constant to be proportional to the Na$^+$ exchange ratio, $\Gamma_{I\_ass} = \Gamma_I \times \gamma_{Na}$, with the prefactor $\Gamma_I$ defined as the frequency at which an ion jumps along the [001] direction in the absence of a defect (**Supplementary Note 2**). This revised model produced excellent agreement with experimental data (**Fig. 3b** and **Supplementary Video 3**). In short, the KMC simulations indicate that the presence of Na$^+$ enhances the inter-channel mobility of Li$^+$, shifting diffusion from 1D to quasi-2D. As $\gamma_{Na}$ increases, the resulting boost in intra-channel diffusivity gives rise to an autocatalytic exchange process, manifesting as superdiffusive behavior.

Furthermore, to confirm the existence of enhanced microscopic cross-channel diffusion, 4D-STEM and in situ synchrotron XRD were conducted at the single-particle and particle-ensemble levels, respectively. **Fig. 3c** summarizes the strain differences between the lithiated or sodiated phases and the olivine FePO$_4$ phase along the [101] ($\varepsilon_{xz}$) and [001] ($\varepsilon_{zz}$) directions, calculated using lattice parameters from standard PDF cards (**Supplementary Table 6**). These two directions were chosen due to their opposite strain responses upon lithiation or sodiation. During lithiation, compressive strains are observed along both [101] and [001], while tensile strains occur for sodiation. As shown in **Supplementary Fig. 13** and **Fig. 3d**, we first set the acquired average lattice parameters of multiple 20nm_FP particles as the reference and then calculated the strain of all particles to exclude potential systematic errors. Notably, those mapped lattice parameters agree with those obtained from XRD refinement (**Supplementary Table 2**). Both 20nm_FP and electrochemically prepared 20nm_Li$_{0.47}$FP exhibit homogeneous $\varepsilon_{xz}$ and $\varepsilon_{zz}$ with narrow distributions and median values close to the theoretically calculated strains (**Fig. 3d-e** and **Extended Fig. 2c**). The uniform compressive strain distribution in 20nm_Li$_{0.47}$FP also indicates a continuous change of lattice parameters, supporting a solid solution transition upon lithiation. Heterogeneity evolved during Li-Na ion exchange (**Figs. 3d-e** and **Extended Fig. 2d**). Interestingly, both enhanced compressive and tensile strains relative to the starting 20nm_Li$_{0.47}$FP are observed within single particles. After 7 hours of soaking, the electrode global composition evolves to Li$_{0.14}$Na$_{0.33}$FePO$_4$. To compare the strain distributions, we introduce quartiles, along with lower whisker (LW) and upper whisker (UW) as indicators, which extend to the smallest and largest data points within 1.5× interquartile range (IQR) from the first (Q1) and third quartiles (Q3), respectively. Data points beyond this range are treated as outliers. Specifically, the Q1 values of $\varepsilon_{zz}$ and $\varepsilon_{xz}$ reach as low as -1.52% and -0.98%, respectively, indicative of higher Li content phases (Li$_X$FePO$_4$, $X > 0.47$) at particle ends (more purplish regions). This provides direct evidence of in-plane migration of structural Li ions within the particles. On the other hand, the UW shifts toward more positive tensile strain values (UW = 0.14% for $\varepsilon_{zz}$ and 1.19% for $\varepsilon_{xz}$) near the long [101] edge of the particles, reflecting increased Na content after Li-Na ion exchange. The local enrichment of Na$^+$ near the long [101] edge facilitates surface wetting and helps relieve volumetric strain penalties during ion exchange.

The conclusions drawn from 4D-STEM are corroborated by tracking phase evolutions of particle aggregates using in situ or ex situ synchrotron XRD (**Fig. 3f-h** and **Extended Fig. 3**). Specifically, we monitored the structural evolution of chemically prepared 20nm_Li$_{0.48}$FP particles during ion exchange in 1M NaCl(aq). As shown in **Fig. 3f** and **Extended Fig. 3a**, as Na continuously replaces Li in the 20nm_Li$_{0.48}$FP, Li redistributes and higher Li-content phases (Li$_X$FePO$_4$, $X > 0.48$) form, evidenced by a leftward shift of the

(020) peak. Since Na preferentially forms $Na_{2/3}FePO_4$ phase and charge neutrality must be maintained, a high vacancy phase (such as FP phase) should form. Li redistribution within solid solution phases requires minimal nucleation; at early times the nucleated high vacancy and $Na_{2/3}FePO_4$ phases occupy low volume fractions, therefore producing an observable left shift of the main phase group. With longer exchange, more Li ions are exchanged out of the structure, and the high vacancy and $Na_{2/3}FePO_4$ phases start to dominate, as further confirmed by ex situ synchrotron XRD after 48 hours soaking (**Fig. 3g** and **Extended Figs. 3a, 3d**). A similar evolution is observed for 20nm_$Li_{0.28}$FP, which has a higher initial vacancy concentration (72%) and achieves deeper exchange, yielding a more dominant FP phase ($Li_{0.03}Na_{0.25}FePO_4$ after 48 hours; **Supplementary Table 7**, and **Extended Figs. 3b-c**). Notably, a high vacancy phase forms only when the initial vacancy concentration is sufficiently high. In contrast, when the initial vacancy concentration is low (e.g., starting from 20nm_$Li_{1.0}$FP with no vacancy), the structure lacks sufficient flexibility to relax into thermodynamically stable phases. Instead, we observe significantly increased in-plane tensile strain and lattice distortion, along with a reduced coherence length, evidenced by a leftward shift and pronounced broadening of the (101) reflection of the initial $LiFePO_4$ phase. This broadened peak spans the full range between the (101) reflections of $NaFePO_4$ and $LiFePO_4$ (**Fig. 3h**).

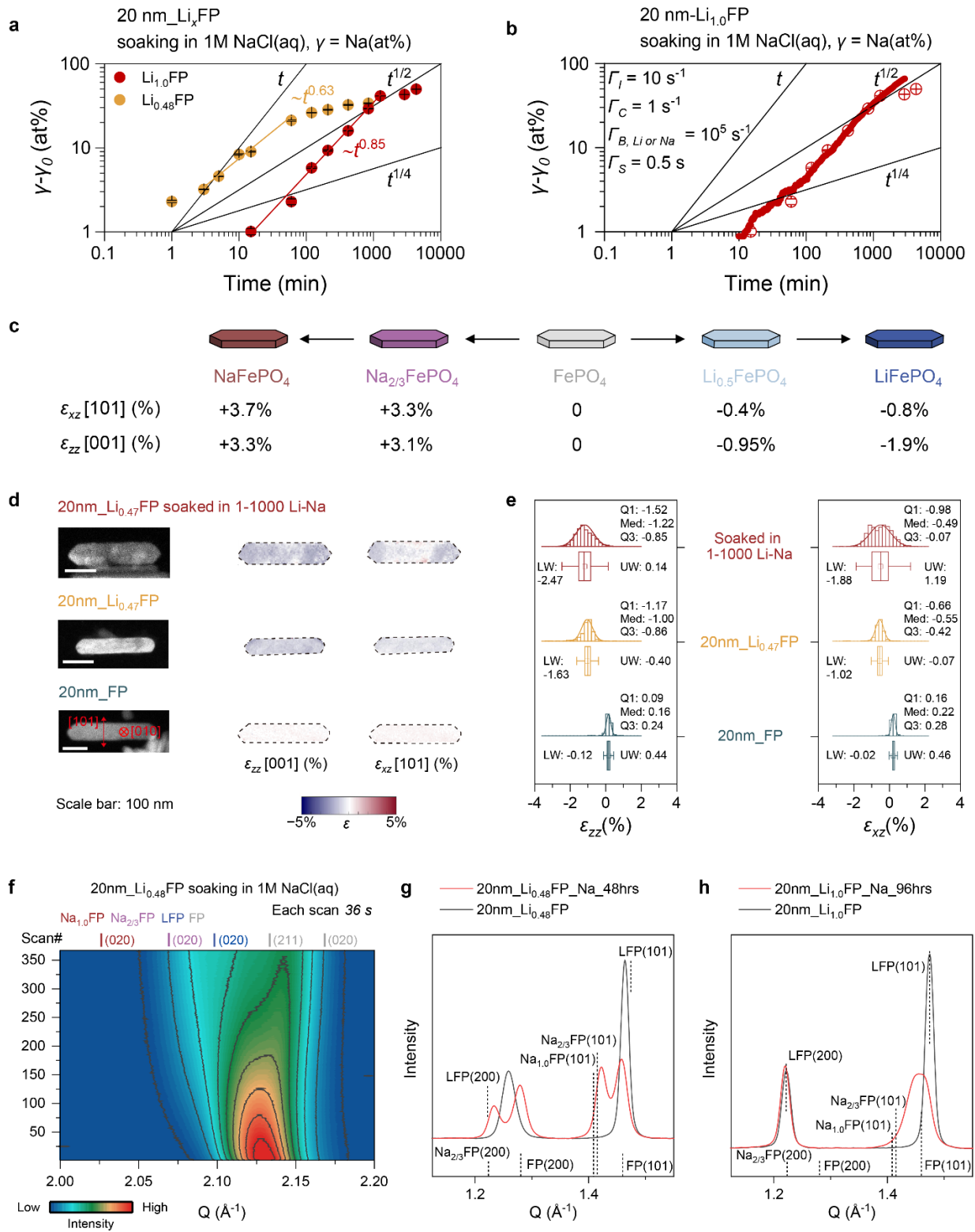

**Fig. 3: Non-Faradaic Li-Na ion exchange. a,** Logarithmic plot of Na tracer atomic ratio changes $(\gamma - \gamma_0)$ over time during Li-Na ion exchange in carbon-coated 20nm_Li$_X$FP ($X$ = 0.48 or 1.0) particles. The solid light orange and red lines are the linear fitting curves, with corresponding fitted time-dependences listed

nearby. **b**, Logarithmic plot of Na tracer atomic ratio changes $(\gamma - \gamma_0)$ over time during Li-Na ion exchange in carbon-coated 20nm_Li$_{1.0}$FP particles (red open circles) with the corresponding KMC simulations (solid red dots). **c**, Theoretical strain difference of all relating phases relative to the olivine FePO$_4$ phase along [101] ($\varepsilon_{xz}$) and [001] ($\varepsilon_{zz}$) directions converted from d-spacing differences (**Supplementary Table 6**). It is worth noting that the lattice parameters of the solid solution Li$_{0.5}$FePO$_4$ phase are calculated according to Vegard's law, which has been experimentally validated for Li$_X$FePO$_4$[63-65]. **d**, STEM images and strain maps of the synthesized 20nm_FP, and electrochemically prepared 20nm_Li$_{0.47}$FP before and after 7 hours soaking in 1mM: 1M LiCl: NaCl(aq) mixture along [001] and [101] directions. Compositions become Li$_{0.14}$Na$_{0.33}$FePO$_4$ after 7 hours soaking. The red arrows in 20nm_FP denote general [101] and [010] orientations of the particles. **e**, Corresponding box plots with a normal distribution summary of strain values from (**d**). In the box plots, the box edges represent the first quartile (Q1, 25$^{th}$ percentile) and third quartile (Q3, 75$^{th}$ percentile), with the interquartile range (IQR) defined as (Q3 − Q1). The line inside the box indicates the median value (Med, 50$^{th}$ percentile), while the square marker denotes the mean value. Lower whisker (LW) and upper whisker (UW) extend to the smallest and largest data points within 1.5×IQR from Q1 and Q3, respectively. Data points beyond this range are treated as outliers. **f**, In situ synchrotron XRD tracking of 20nm_Li$_{0.48}$FP particles during ion exchange in 1M NaCl(aq). **g**, Ex situ synchrotron XRD of 20nm_Li$_{0.48}$FP before and after 48 hours soaking in 1M NaCl(aq). **h**, Ex situ synchrotron XRD of 20nm_Li$_{1.0}$FP before and after 96 hours soaking in 1M NaCl(aq). Error bars in (**a-b**) represent the standard deviation of three replicate measurements.

Based on KMC simulations, 4D-STEM, and synchrotron XRD results, we illustrated the effects of Na$^+$ on diffusion dynamics in **Fig. 4a-b**. Microscopically, since the starting Li phase is either in solid solution or full, all the channels are filled with Li ions. During ion exchange, Na$^+$ from the solution replaces structural Li$^+$, leading to the formation of tapered channels, with one end favoring Li and the other favoring Na. Our previous computational results also indicate that Li and Na prefer phase separation when co-intercalated into the same channel[64]. The development of tapered channels and the thermodynamic tendency for phase separation contribute to directional motion. Additionally, Na$^+$ induces enhanced cross-channel diffusion of Li$^+$ due to the increased in-plane tensile strain and lattice distortion, shifting the diffusion dimensionality from 1D to quasi-2D. We also emphasize that the observed apparent superdiffusion does not stem from a heavy-tailed jump-length distribution or strict adherence to Lévy walk behavior. Instead, it can be interpreted within a coupled reaction-diffusion framework, where surface exchange serves as the rate-limiting step. The extreme chemical potential difference between structural Li$^+$ and Li$^+$ in solution ($\mu_{Li^+} \sim \log(0) \to -\infty$) drives "delithiation" and subsequent Na$^+$ exchange. The system thus operates in a surface-reaction-limited regime, exhibiting an exponential exchange profile that can be described as $(\gamma - \gamma_0) \sim (1 - \exp(-t/\tau))$, where $\tau$ denotes the characteristic exchange time (**Supplementary Note 1** and **Supplementary Fig. 14**).

By contrast, the 340 nm microplatelet particles, as shown in **Fig. 4c**, only exhibit close to normal diffusion behavior ($\alpha/2 \approx 0.46$), in contrast to the apparent superdiffusive behavior observed in 20 nm and 45 nm nanoplatelets. Compared to 20nm_Li$_{0.28}$FP with similar Li concentration, the ion exchange in 340nm_Li$_{0.24}$FP takes significantly longer to reach equilibrium (~4 days vs. ~7 hrs) and results in much less Na being exchanged (~16% vs. ~90%) (**Extended Fig. 2a**). Notably, the 340nm_Li$_{0.24}$FP particles show minimal changes in the XRD peak profile after Li-Na ion exchange, with only a slight broadening of the LFP (020) peak (**Fig. 4d**). This is opposite to the continuous lattice evolutions with new phase groups emerging in 20nm_Li$_{0.28}$FP particles (**Extended Fig. 3c**). These observations suggest that nanoplatelet particles have a more flexible lattice structure to accommodate the volumetric strain induced by Na$^+$. In contrast, microplatelet particles cannot resolve this strain, suppressing superdiffusion and displaying only the 'surface wetting' effect[46,47,59] with a shallow exchange degree and ordinary diffusion behavior.

The more detailed atomic-level XAS analysis further confirms the flexible lattice framework of the nanoplatelet particles. As expected, both 20 nm and 340 nm particles show a transition from Fe$^{III}$ to Fe$^{II}$ with increased lithiation degree, indicated by the left shift in the absorption edge energies in X-ray absorption near-edge spectroscopy (XANES) analysis (**Extended Figs. 4a-b**). One would also expect the reduction of Fe$^{3+}$ to Fe$^{2+}$ to increase the average Fe-O bond length in both cases. However, differences in the evolution of the Fe-O scattering path become apparent between the two particles when analyzing the Fourier transform of the extended X-ray absorption fine structures (FT-EXAFS) and wavelet transforms of the weighted EXAFS signals (**Figs. 4e-f**, and **Extended Figs. 4c-h**). 20 nm nanoplatelets exhibit a single Fe-O radial distance during lithiation, with an average Fe-O bond length of 2.02 Å at 47% lithiation. The more transiently equivalent Fe-O bond lengths in the [Fe$^{II}$O$_6$] and [Fe$^{III}$O$_6$] sites facilitate concerted electron hopping across a larger lattice domain (**Fig. 4g**)[66]. In contrast, 340 nm particles show two distinct groups of Fe-O bond upon lithiation: one around 1.94 Å and another around 2.14 Å at a 45% lithiation degree (**Supplementary Tables 8 and 9**). Notably, lattice softness is closely linked to the Debye-Waller factor ($\sigma^2$), which reflects the impact of atomic vibrations on X-ray or neutron scattering intensities. A larger $\sigma^2$ indicates a softer lattice due to more pronounced thermal vibrations[67]. In 20nm_Li$_{0.47}$FP ($\sigma^2$ = 0.0104 Å$^2$), the Debye-Waller factor for the Fe-O scattering path is nearly an order of magnitude higher than that in 340nm_Li$_{0.45}$FP (For both Fe-O groups, $\sigma^2$ = 0.0014 Å$^2$), underscoring the increased lattice softness in the smaller particles. In other words, the softer phosphate lattice in 20 nm nanoplatelets supports beneficial vibrational phonon modes that enhance polaron transport.

Additional evidence for the critical role of the lattice framework in diffusion dynamics is provided by $^{57}$Fe Mössbauer spectroscopy, which is highly sensitive to the oxidation state, electron configuration, and local coordination environment of Fe nuclei. As shown in **Extended Figs. 5a-b** and **Supplementary Fig. 15**, we obtained spectra for both 20 nm nanoplatelet and 340 nm microplatelet at various lithiation degrees to compare the local structures of the [FeO$_6$] sites in the lattice. All spectra were fitted with two doublets assigned to Fe in [Fe$^{II}$O$_6$] or [Fe$^{III}$O$_6$], with the fitted parameters summarized in **Supplementary Tables 10 and 11**. As further summarized in **Extended Fig. 5c**, quadrupole splitting ($\Delta E_Q$) for 340 nm particles exhibits minimal changes for both Fe$^{II}$ and Fe$^{III}$ at different lithiation degrees. However, for 20 nm particles, both $\Delta E_Q$-Fe(II) and $\Delta E_Q$-Fe(III) decrease in partially lithiated states compared to the empty or fully lithiated phases (20nm_FP and 20nm_Li$_{1.0}$FP). For instance, in 20nm_Li$_{0.48}$FP, $\Delta E_Q$-Fe(II) and $\Delta E_Q$-Fe(III) are 2.81 mm/s and 1.31 mm/s, respectively, while in 20nm_Li$_{1.0}$FP, $\Delta E_Q$-Fe(II) is 3.1 mm/s and in 20nm_FP, $\Delta E_Q$-Fe(III) is 1.49 mm/s (**Extended Fig. 5**). This suggests a more symmetric electron distribution around the Fe nucleus in 20 nm particles during the transition, indicating a decreased lattice polarizability and a larger degree of electron/Li delocalization. These findings also align with the more equivalent Fe-O bond lengths in the $[Fe^{II}O_6]$ and $[Fe^{III}O_6]$ sites observed in the XAS analysis of 20 nm particles.

Overall, both XAS analysis and Mössbauer spectroscopy reveal distinct lattice softness in 20 nm and 340 nm $Li_XFePO_4$ particles, despite their chemical identity. The more flexible and less polarized lattice of the 20 nm particles facilitates concerted polaron hopping, contributing to $Na^+$-induced quasi-2D diffusion. Given the absence of external electron flow, it may seem counterintuitive that polaron transport is crucial for the non-Faradaic Li-Na ion exchange. However, it is essential to consider the strong bond between the $Li^+/Na^+$ in the structure and the localized electron on the Fe. An electron redistribution needs to occur within the particles during the ion exchange and ion hopping, since new Li and Na phases with varying degrees of lithiation and sodiation need to evolve.

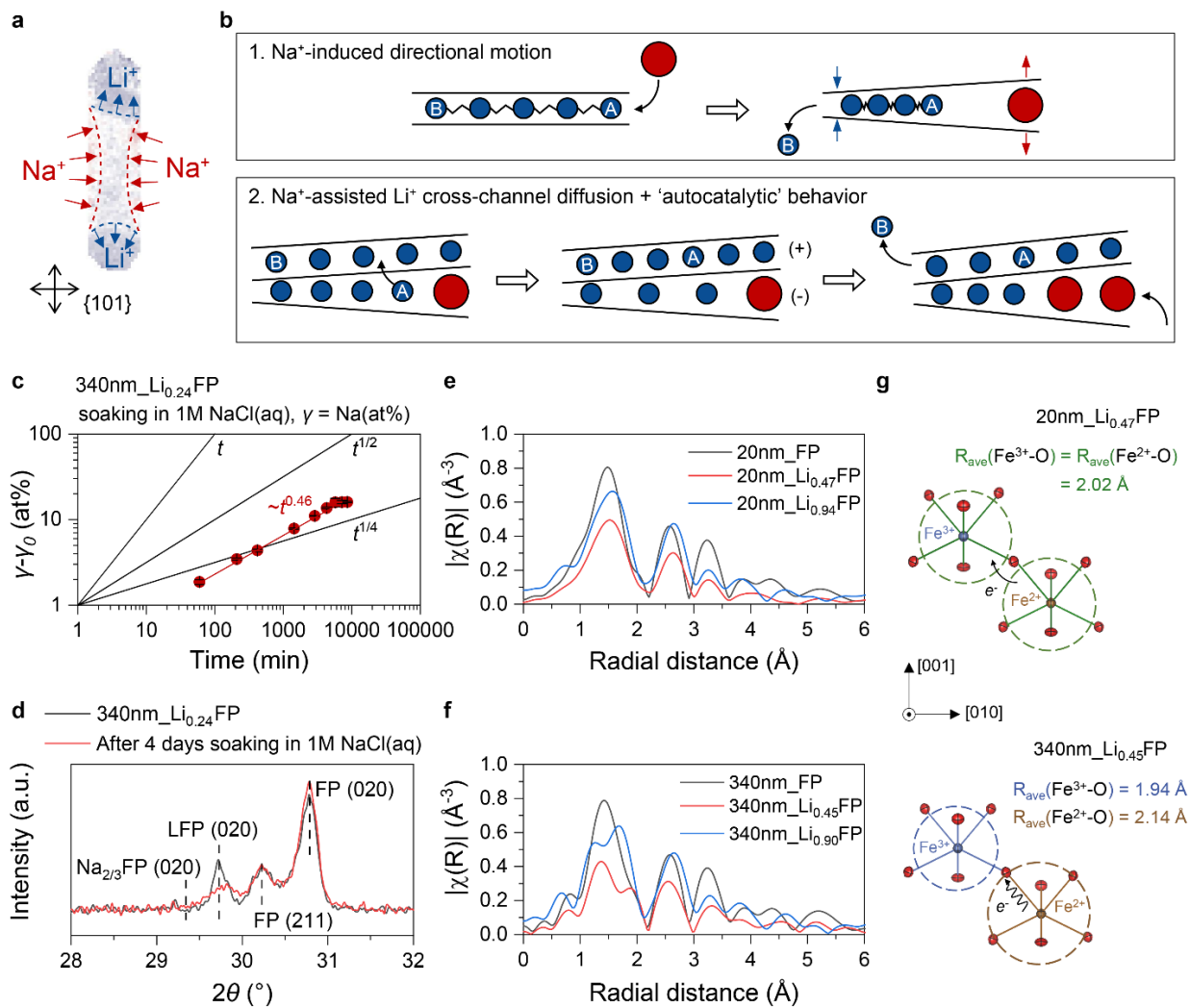

**Fig. 4: Effects from Na ions and lattice vibrations on diffusion behavior. a**, Strain distribution of the soaked 20nm_Li$_{0.47}$FP particle along the [101] direction, with arrows indicating the diffusion fronts of Na and Li ions. Na ions diffuse into the structure from the long edge, while the structural Li ions show enhanced cross-channel diffusion, concentrating near the ends of the particle. **b**, Schematic illustrating the effects of Na$^+$: 1. Na$^+$-induced directional motion; 2. Na$^+$-assisted Li$^+$ cross-channel diffusion. Key: Li ions (navy); Na ions (maroon). **c**, Logarithmic plot of Na tracer atomic ratio change $(\gamma - \gamma_0)$ over time during Li-Na ion exchange in carbon-coated 340nm_Li$_{0.24}$FP particles. The solid red line is the linear fitting curve, with the corresponding fitted time-dependency. **d**, *In-house* XRD of 340nm_Li$_{0.24}$FP before and after 4 days soaking in 1M NaCl(aq). **e**, Fourier-transformed EXAFS in real-space (R-space) of 20nm_FP/Li$_{0.47}$FP/Li$_{0.94}$FP. **f**, Fourier-transformed EXAFS in real-space (R-space) of 340nm_FP/Li$_{0.45}$FP/Li$_{0.90}$FP. **g**, Schematic illustrating more equivalent Fe-O bond lengths in the [Fe$^{II}$O$_6$] and [Fe$^{III}$O$_6$] sites for nanoplatelet particles than those of microplatelet particles, facilitating electron hopping.
18

**Conclusions**

Classical Fickian diffusion models, built on the assumption of uncorrelated, memoryless, and non-excluding motion, have long dominated the interpretation of transport phenomena in electrochemical systems. However, this framework breaks down in solid-state ionic conductors, where strong steric constraints prevent ion bypassing, giving rise to complex, non-Fickian dynamics. Despite its significance, this regime remains underexplored, even in well-studied materials like $LiFePO_4$.

In this study, we uncover pronounced non-Fickian ion diffusion during Li ion exchange with both Li isotopes and Na ions in $Li_X FePO_4$ using direct tracer exchange measurements, which decouple ionic from electronic contributions. Isotope exchange confirms SFD behavior, whereas Li-Na exchange reveals an autocatalytic rate increase and an apparent superdiffusive response which may guide further analysis considering Maxwell-Stefan couplings for the $Li^+/Na^+$ and the vacancies[68]. Tracer exchange measurements further allow us to probe the intricate kinetics and disentangle the effects of surface exchange, bulk diffusivity, and lattice distortions. Supported by KMC simulations, our analysis extracts key material parameters, such as surface exchange rates and cross-channel hopping frequencies. Altogether, this methodology provides a powerful framework for understanding and controlling ionic diffusion in solids. Advanced characterization further shows that particle size and carbon coating play unexpected roles in modulating ion diffusion. Nanosizing not only reduces diffusion length and increases surface-to-volume ratio, but also enhances lattice flexibility and electron transport.

Overall, this work presents a unified strategy to probe non-Fickian transport in confined solids. Using a simple and generalizable tracer-exchange protocol, we circumvent the limitations of standard electrochemical methods. Our findings highlight the rich interplay among surface exchange, lattice structure, transport dimensionality, and electron-ion coupling, offering both conceptual and methodological tools to guide the design of next-generation materials for energy storage, ion separation, and solid-state ionics.

**Methods**

**Preparation of 1 M $^6$LiCl(aq) solution.** First, 150 ml of 2 M $^6$LiOH(aq) was prepared as follows: 0.3 mol of $^6$Li metal (Cambridge Isotope Laboratories, 95 at% $^6$Li, 5 at% $^7$Li) was brushed and weighed inside a glovebox to prevent oxidation. The $^6$Li metal was then quickly transferred into refrigerated deionized (DI) water (4°C) in an ice bath to generate $^6$LiOH(aq). The resulting solution was then transferred into a 150 ml volumetric flask, and the volume was adjusted with additional DI water. Concentrated HCl(aq) (Sigma-Aldrich, 37 wt% HCl) was then added dropwise to the prepared 2 M LiOH(aq) in an ice bath until a neutral pH was reached. The final mixture was transferred to a 300 ml volumetric flask, and the volume was adjusted with additional DI water, yielding 1 M $^6$LiCl(aq). The composition was verified with ICP-MS (**Supplementary Table 1**), confirming a $^6$Li abundance of 94.8 ± 0.1 at%, consistent with the provided purity.

**Synthesis of carbon-coated 20nm/45nm/340nm_Li$_{1.0}$FP particles.** All three pristine LiFePO$_4$ particles with different diffusion channel lengths were synthesized using a solvothermal method, following our previously reported recipes[45]. After the solvothermal synthesis was completed, all pristine LiFePO$_4$ particles followed the same washing, carbon coating process described in the following to prepare the carbon-coated LiFePO$_4$ particles. Specifically, the obtained LiFePO$_4$ precipitates from the solvothermal synthesis were centrifuged three times with deionized water and ethanol, followed by 60 °C drying overnight. To further increase the electronic conductivity of LiFePO$_4$ particles, surface carbon coating is utilized, which has proven to be an effective strategy. Concretely, the carbon coating procedure involved amalgamating pristine LiFePO$_4$ with sucrose (as the carbon source) in a mass ratio of 5: 1 (LiFePO$_4$: sucrose), while preserving the integrity of the primary particles. The mixture was first calcinated in an Ar atmosphere at 200 °C for 0.5 h, followed by heating to 550 °C for 2.5 h. The heating rate was 3 °C min$^{-1}$.

**Synthesis of annealed bare 20nm_Li$_{1.0}$FP particles.** For the preparation of 20nm_Li$_{1.0}$FP particles without carbon coating, the pristine 20nm_Li$_{1.0}$FP particles were annealed under the same condition described above without mixing with sucrose.

**Synthesis of chemically prepared partially lithiated 20nm_Li$_{0.28/0.48/0.76}$FP and 340nm_Li$_{0.24/0.32/0.64}$FP particles.** Carbon-coated FePO$_4$ particles were first prepared by fully extracting Li from carbon-coated LiFePO$_4$ particles. Specifically, an oxidizing solution was made by dissolving 0.85 g of nitronium tetrafluoroborate (NO$_2$BF$_4$) in 50 mL of acetonitrile. Then, 0.5 g of carbon-coated LiFePO$_4$ powder was immersed in the solution and stirred for 24 hours at room temperature (20-25 °C). The powder was washed several times with acetonitrile and dried in a vacuum oven at 80 °C for 12 hours. The resulting FePO$_4$ particles were verified by XRD (**Supplementary Figs. 4f-h** and **Supplementary Table 2**). Partially lithiated particles were synthesized by reacting LiI and FePO$_4$ in acetonitrile for 24 hours in a glovebox. Specifically, 0.3, 0.5, or 0.8 units of LiI per unit of FePO$_4$ were suspended in acetonitrile and stirred for 24 hours, with a LiI concentration of 1 M. The resulting powder was washed several times with acetonitrile and dried in a vacuum oven at 80 °C for 12 hours[69]. The final lithium concentrations of the particles were verified by ICP-MS (**Supplementary Table 5**).

**Preparation of electrodes.** All carbon-coated Li$_X$FePO$_4$ electrodes were prepared by casting a slurry of Li$_X$FePO$_4$, Super P carbon black (MTI Corporation; Item Number: Lib-SP; average particle size ~40 nm; purity ≥ 99.5%), and polyvinylidene fluoride (MTI Corporation; Item Number: Lib-PVDF; purity ≥ 99.5%) with a mass ratio of 80:10:10, in N-methyl-2-pyrrolidone. The working electrodes (~2.5 mg/cm$^2$) were prepared by drop casting the slurry on a 0.5 × 0.5 cm$^2$ geometrical surface of a carbon paper (TGP-H-060, Fuel Cell Earth, 190 μm in thickness, 78% porosity) current collector of 2.5 × 0.5 cm$^2$. Besides, to increase the hydrophilicity of the carbon paper, the 0.5 × 0.5 cm$^2$ drop-casting area was cleaned with argon plasma at 100 watts for 1 minute before the drop-casting of the slurry. The preparation of bare 20nm_Li$_{1.0}$FP electrodes follows a slightly different recipe reported previously[70]. Specifically, 30 wt% active material (annealed bare 20nm_L$_{1.0}$FP), 65% wt% carbon black and 5 wt% binder are used to prepare slurry, with an average mass loading of ~200 ug. FePO$_4$ counter electrodes were made with the slurry using the same mass ratios depositing on carbon felt (Alfa Aesar) disks (0.9525 cm diameter × 3.18 mm thickness, around 240 g/m$^2$ in areal weight). The active material mass loading on the counter electrodes ranged between 60 and 70 mg cm$^{-2}$. Li$_X$FePO$_4$/Na$_Y$FePO$_4$ counter electrodes were prepared electrochemically. Specifically, FePO$_4$ counter electrodes were galvanostatic lithiation/sodiation in 1M LiCl(aq)/NaCl(aq) at a C/20 (8.5 mA/g) rate until reaching a -0.6 V versus Ag/AgCl voltage cutoff. The larger mass loading of the counter electrode ensures we have enough ion stock in the counter electrode to avoid side reactions from water splitting or pH fluctuations. C/N describes the current to (de)intercalate the electrode in Nh.

**Electrochemical methods.** All electrochemical operations were performed on a Bio-Logic VMP3 workstation using a three-neck round-bottomed flask at room temperature (20 ~ 25 °C). N$_2$ (purity > 99.998%) was continuously bubbled into the solution to avoid side reactions caused by dissolved O$_2$[45,64,71]. To verify the quality and measure the accessible capacity of fabricated working electrodes, the working electrodes were cycled in 60 mL 1 M LiCl aqueous solutions (17 mA/g; paired with Li$_X$FePO$_4$ counter electrodes) between -0.6 V and 0.6 V (vs. Ag/AgCl (sat. KCl)) at room temperature (20 ~ 25 °C) (**Supplementary Fig. 5**).

**Chronoamperometry experiments.** Before each voltage step, the working electrodes were discharged at C/10 to -0.6V (vs. Ag/AgCl (sat. KCl)) and allowed to relax for 4 hours to reach equilibrium. Three different overpotentials (η) were selected for the voltage steps: 0.025 V, 0.2 V, and 0.4 V. The reference voltage was determined by averaging the voltage plateaus of charge and discharge during the 17 mA/g cycle of both the coated and uncoated samples. A cutoff current of C/50 was applied for each step.

**X-ray diffraction (XRD) characterization.** For in-house measurements of synthesized LiFePO$_4$ and FePO$_4$ powder, XRD was carried out on Rigaku MiniFlex 600 diffractometer, using Cu Kα radiation (Kα1: 1.54059 Å; Kα2: 1.54441 Å; Kα12 ratio: 0.4970). The tube voltage and the current used were 40 kV and 15 mA. Diffractograms were repeated three times to increase the S/N ratio with a 0.02° step width and a 10°/min speed. Rietveld refinement was executed using GSAS-II software (**Supplementary Figs. 4 and 11**, and **Supplementary Tables 2 and 4**). For in-house measurements of carbon cloth or carbon paper electrodes, XRD was carried out on Rigaku SmartLab multipurpose diffractometer, using Cu Kα radiation (Kα 1: 1.54059 Å; Kα 2: 1.54441 Å; Kα 12 ratio: 0.4970). The tube voltage and the current used were 40 kV and 40 mA. Diffractograms were repeated five times to increase the S/N ratio with a 0.02° step width and a 10°/min speed. In situ and ex situ synchrotron XRD measurements were conducted at 33-BM[72] beamlines of Advanced Photon Source (APS) and the Stanford Synchrotron Radiation Lightsource (SSRL) 11-3 beamline. A specially designed three-electrode cell was used for in situ measurements, allowing aqueous electrolyte solution to flow across the electrode while changing the current and monitoring the phase transformation by synchrotron. During the lithiation of the electrodes, 1 M LiCl(aq) was used as electrolytes, paired with Li$_X$FePO$_4$ carbon felt counter electrodes and one leakless miniature Ag/AgCl reference electrode (Edaq Inc, ET072-1). During non-Faradaic ion exchange of partially lithiated particle, 1 M NaCl(aq) was used as electrolytes, paired with Na$_Y$FePO$_4$ carbon felt counter electrodes and leakless miniature Ag/AgCl reference electrode (Edaq Inc, ET072-1).

**4D-Scanning transmission electron microscopy (4D-STEM) characterization.** For STEM characterization, the FePO$_4$ nanoplates were first dispersed and sonicated in isopropyl alcohol (MACRON). 5 μL of the dispersion was then drop-casted on a TEM grid (CF400-Cu, Electron Microscopy Sciences) and dried in the air for 24 h. The grid was plasma cleaned (Tergeo EM Plasma Cleaner, PIE Scientific) to remove carbon contamination before STEM characterization. 4D-STEM was performed on a Thermo Talos F200X TEM-STEM operated at 200 kV. The microscope was operated in the μProbe STEM mode with a semi-convergence angle of 0.46 mrad, a 20 μm condenser aperture (C2), and a camera length of 205 mm. The corresponding probe size is 1.8 nm in full width at half-maximum. A double-tilt holder was used to choose the desired crystallographic axis ([010]) for 4D-STEM data collection. The electron probe was then raster-scanned across the particles at a step size of 2 nm, with a diffraction pattern recorded at each probe position using a CMOS camera (Ceta, Thermo Scientific). The dwell time was 20 ms for each step. The calculation of strain from 4D-STEM data is elaborated in **Supplementary Note 6**.

**Inductively coupled plasma-mass spectrometry (ICP-MS) characterization.** 3% HNO$_3$(aq) was used as the diluting matrix for all the Li recovery solutions. The non-Faradaic ion exchanged particles were first washed 3-5 times with 60 ml fresh distilled water and then digested with an aqua regia solution for three days to ensure complete dissolution. The resulting supernatant was diluted with 3% HNO$_3$ for later ICP-MS measurement. All the measurements used either Thermo iCAP Q ICP-MS or Thermo iCAP RQ ICP-MS.

**Scanning electron microscopy (SEM) characterization.** Scanning electron microscopy (SEM, Zeiss Merlin) was performed at the accelerating voltage of 10 kV.

**X-ray absorption spectroscopy (XAS) analysis.** XAS measurements were carried out at the SSRL 11-2 beamline. The measurements at the Fe K-edge were performed in transmission mode. All XAS data analysis were performed with Athena software package to extract XANES and EXAFS. Fourier Transform of Fe K-edge EXAFS was performed by using Hanning window function with k-weights of 2. For model-based EXAFS analysis, all the scattering paths were generated by the FEFF calculation function in Artemis based on the crystal structure of LiFePO$_4$. All those EXAFS fittings are done with k-range from 3 to 10.5 Å$^{-1}$ to estimate scattering path length. Wavelet Transform were done by HAMA FORTRAN VERSION with Morlet wavelet (k-range from 3 to 10.5 Å$^{-1}$ and R-range from 1 to 2.5 Å).

**Mössbauer spectroscopy.** Conventional Mössbauer data were collected at the APS. The instrument is equipped with a $^{57}$Co/Pd source in constant acceleration mode, and a Vortex silicon drift detector with a resolution of 150 eV at 14.4 keV. The spectrum collected at ambient temperature from a 3 μm $^{57}$Fe enriched iron foil was used for calibration. Samples were contained in an aluminum foil with a 6 mm hole and taped to a lead foil with a 5 mm hole to suppress background. Data was collected in transmission geometry. All data were collected at ambient conditions and processed using WinNormos (Wissel Co.).

**Data availability**

The data used in this study are available in the main text and the Supplementary Information. All other data are available from the corresponding author upon request.

**Acknowledgments**

We thank Reginaldo J. Gomes (University of Chicago) from Prof. Chibueze Amanchukwu's group for the providing of 2 M $^6$LiOH(aq). We thank Evguenia Karapetrova from Argonne National Laboratory for assistance with synchrotron XRD measurements at 33BM-beamline of APS. We thank Vivek Thampy, Molleigh B. Preefer and Christopher J. Takacs from SLAC National Accelerator Laboratory for setting up synchrotron XRD measurements at SSRL 11-3 beamline. G.Y. and C.L acknowledge support from the Energy Storage Research Alliance (ESRA, DE-AC02-06CH11357), an Energy Innovation Hub funded by the U.S. Department of Energy (DOE) Office of Science, Basic Energy Sciences (BES), and this work made use of instruments in the Electron Microscopy Core, Research Resources Center in University of Illinois at Chicago. P.O. and M.W. acknowledge support from Shell Global Solutions International B.V. Z.T. and Q.C. acknowledge support from U.S. DOE, Office of Science, BES under Award No. DE-SC0024064 for 4D-STEM data collection and analysis. M.W. and H.Z. acknowledge use of the SSRL, SLAC National Accelerator Laboratory, supported by the U.S. DOE, Office of Science, BES under Contract No. DE-AC02-76SF00515. B.L. and E.E.A. acknowledge use of the APS, a U.S. DOE Office of Science User Facility operated for the DOE Office of Science by Argonne National Laboratory under Contract No. DE-AC02-06CH11357, and B.L. further acknowledges support from the National Science Foundation, Division of Earth Sciences (EAR) through SEES (EAR –2223273).

**Author contributions**

G.Y. conceived the idea, led all aspects of the experimental investigation, contributed to the simulation analysis and diffusion dynamics derivations, and coordinated the project. P.O. performed and interpreted the KMC simulations, contributed to the diffusion dynamics derivations, and co-coordinated the project. Z.T. carried out 4D-STEM characterization. S.P. provided insights into diffusion dynamics derivations and CIET theory. M.W. and H.Z. conducted XAS characterization. B.L. and E.E.A. conducted Mössbauer spectroscopy. A.V. assisted with the KMC simulations. S.C., M.W., D.Y. and Q.L. contributed to experimental data collection and analysis. M.Z.B. supervised the CIET theory investigation and diffusion dynamics derivations. Q.C. supervised the 4D-STEM characterization. M.W. supervised the KMC simulations. C.L. supervised the experimental components of the work. G.Y. and P.O. wrote the manuscript with input from all co-authors.

## Competing interests
The authors declare no competing interests.

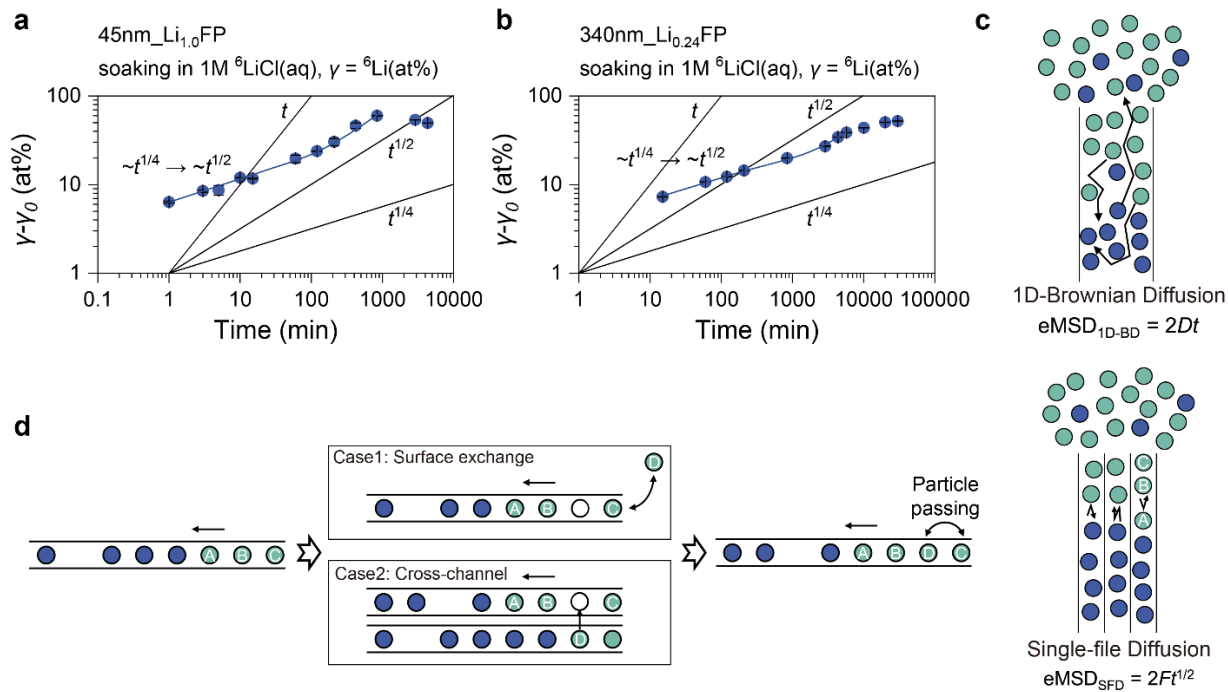

**Extended Fig. 1 More subdiffusion behavior observed and the microscopic illustration of ion passing events. a**, Logarithmic plot of $^6$Li tracer atomic ratio changes $(\gamma - \gamma_0)$ over time during $^6$Li-$^7$Li self diffusion in carbon-coated 45nm_Li$_{1.0}$FP particles. The solid blue line is to guide the eye. **b**, Logarithmic plot of $^6$Li tracer atomic ratio changes $(\gamma - \gamma_0)$ over time during $^6$Li-$^7$Li self diffusion in carbon-coated 340nm_Li$_{0.24}$FP particles. The solid blue line is to guide the eye. **c**, Cartoon in the top panel showing 1D Brownian Diffusion (1D-BD) dynamics during ion exchange between $^7$Li ions (dark blue) and $^6$Li ions (light green). The hypothetical channel is larger than the lithium ions. Although macroscopic diffusion is unidirectional, the microscopic motion of each ion is 3D. In other words, the particles can bypass each other, leading to an ordinary diffusion behavior. Cartoon in the bottom panel showing SFD dynamics during ion exchange between $^7$Li ions and $^6$Li ions. The [010] diffusion channel is only slightly larger than the lithium ions, constraining the Li$^+$ from bypassing each other and leading to the preservation of initial order. **d**, Schematics showing potential atomistic origins of ion passage events due to surface exchange or cross-channel diffusion. Error bars in (**a-b**) represent the standard deviation of three replicate measurements.

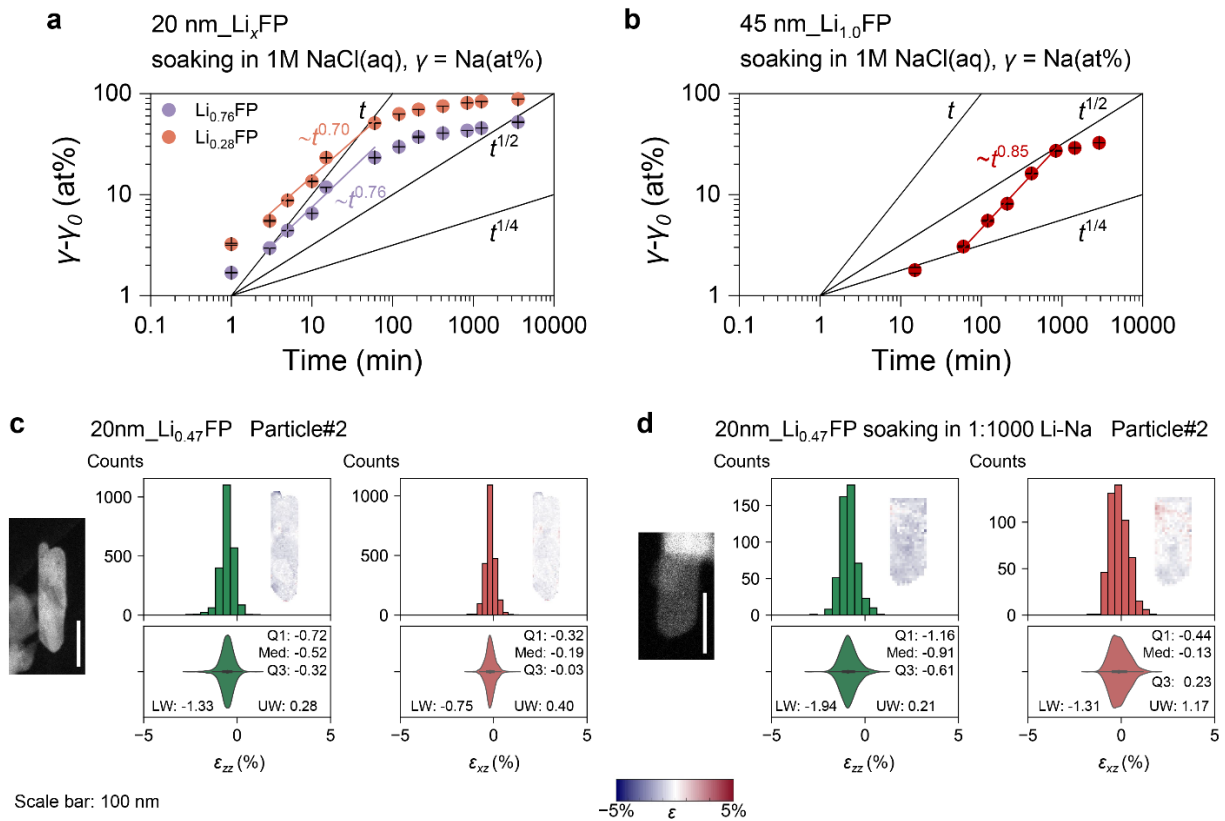

**Extended Fig. 2 More superdiffusion behavior observed and more particles analyzed by 4D-STEM.**
**a**, Logarithmic plot of Na tracer atomic ratio changes $(\gamma - \gamma_0)$ over time during Li-Na ion exchange in carbon-coated 20nm_Li$_X$FP ($X$ = 0.28 or 0.76) particles (See **Methods** for more synthesis details). The solid orange and purple lines are linear fitting curves, with corresponding fitted time-dependences. **b**, Logarithmic plot of Na tracer atomic ratio changes $(\gamma - \gamma_0)$ over time during Li-Na ion exchange in carbon-coated 45nm_Li$_{1.0}$FP particles. The solid red line is the linear fitting curve, with the corresponding fitted time-dependency. **c-d**, Strain maps along [001] and [101] directions with summary statistics for additional 20nm_Li$_{0.47}$FP particles before (**c**) and after 7 hours soaking in 1mM: 1M LiCl: NaCl(aq) mixture (**d**). In the violin plots, first quartile (Q1, 25[th] percentile), median (Med, 50[th] percentile), and third quartile (Q3, 75[th] percentile) are presented, with the interquartile range (IQR) defined as (Q3 − Q1). The lower whisker (LW) and upper whisker (UW) extend to the smallest and largest data points within 1.5×IQR from Q1 and Q3, respectively, while data points beyond this range are treated as outliers. Error bars in (**a-b**) represent the standard deviation of three replicate measurements.

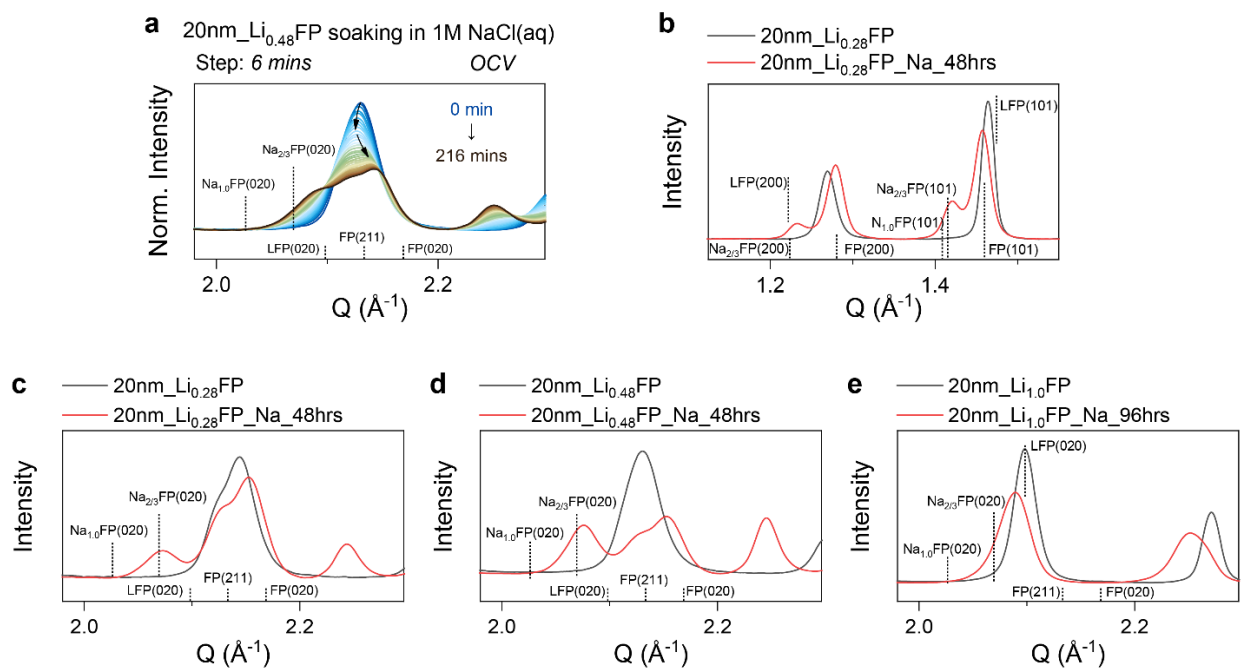

**Extended Fig. 3 Ex situ synchrotron XRD analysis of 20nm_Li$_X$FP particles before and after ion exchange in 1M NaCl(aq). a**, Snapshots of in situ synchrotron XRD of 20nm_Li$_{0.48}$FP particles during ion exchange in 1M NaCl(aq). **b-c**, 20nm_Li$_{0.28}$FP before and after 48 hours soaking in 1M NaCl(aq). **d**, 20nm_Li$_{0.48}$FP before and after 48 hours soaking in 1M NaCl(aq). **e**, 20nm_Li$_{1.0}$FP before and after 96 hours soaking in 1M NaCl(aq).

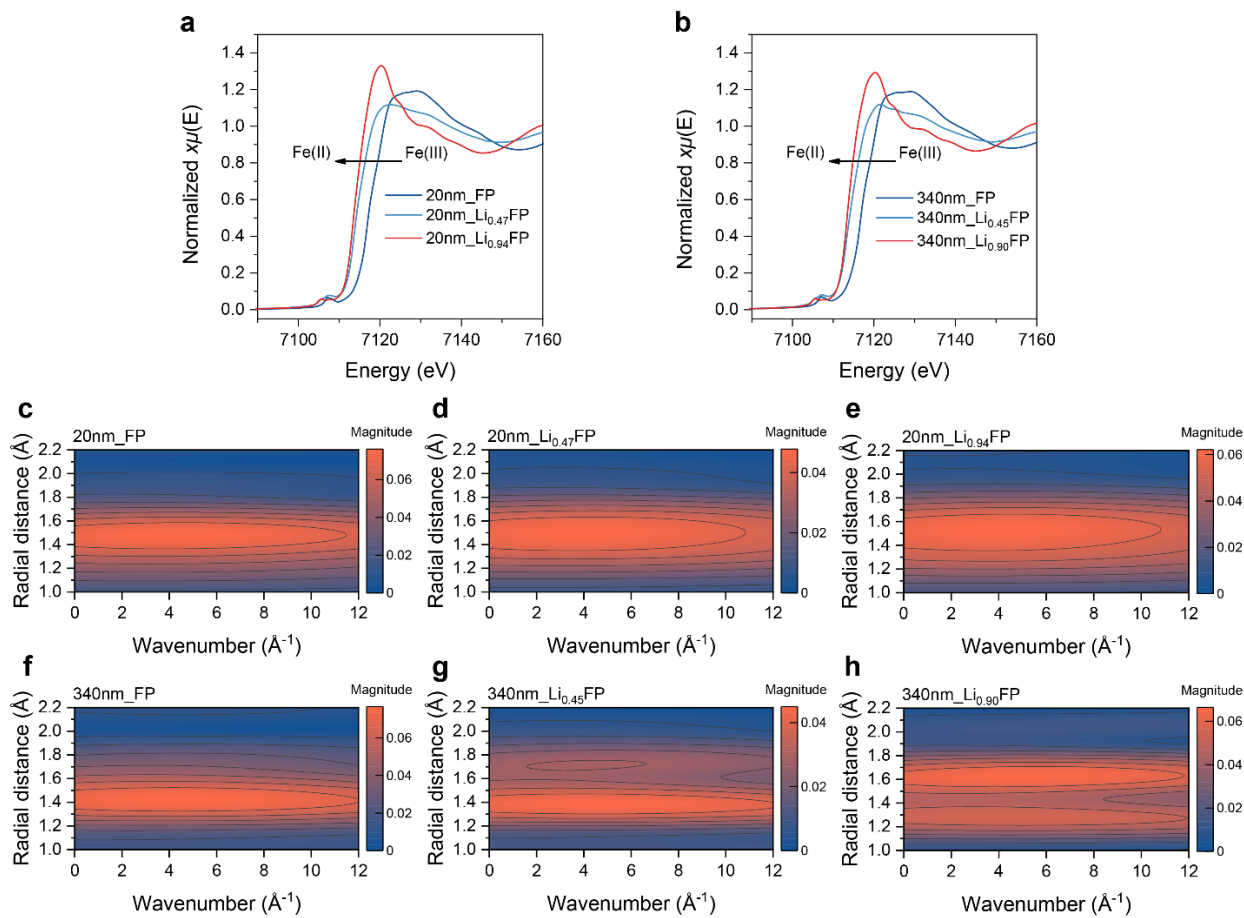

**Extended Fig. 4 XAS analysis. a-b**, Fe K-edge XANES spectra of 20nm_FP/Li$_{0.47}$FP/Li$_{0.94}$FP (**a**) and 340nm_FP/Li$_{0.45}$FP/Li$_{0.90}$FP (**b**). **c-h**, Wavelet transforms for the weighted EXAFS signals of 20nm_FP (**c**), 20nm_Li$_{0.47}$FP (**d**), 20nm_Li$_{0.94}$FP (**e**), 340nm_FP (**f**), 340nm_Li$_{0.45}$FP (**g**), and 340nm_Li$_{0.90}$FP (**h**).

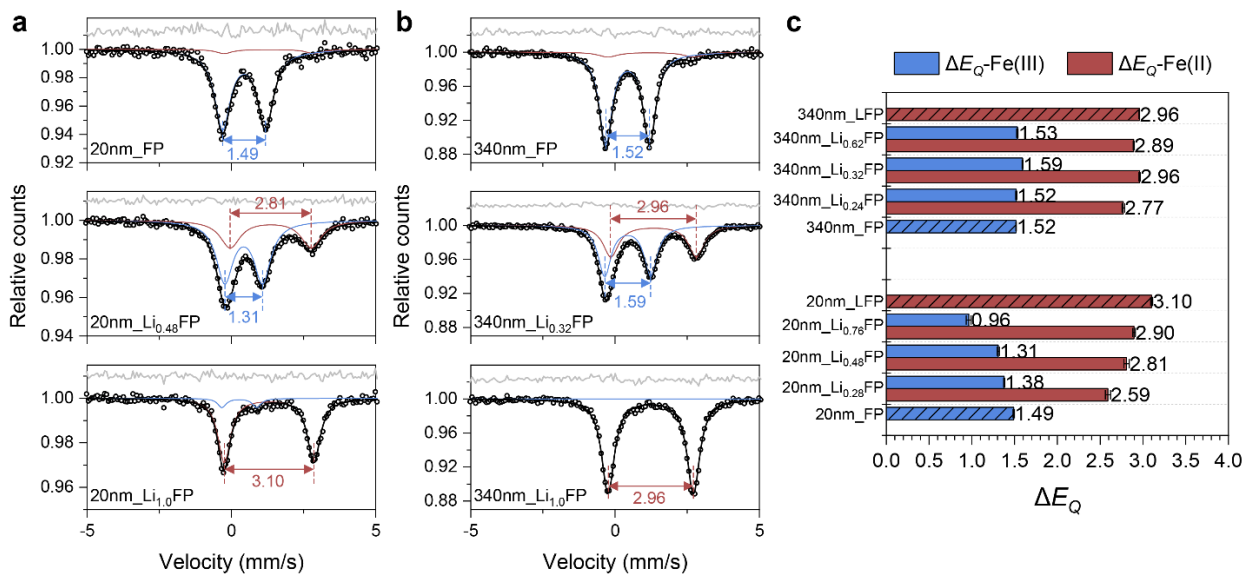

**Extended Fig. 5 Analysis of Fe coordination environment. a**, The $^{57}$Fe Mössbauer spectra of 20nm_FP/Li$_{0.48}$FP/Li$_{1.0}$FP with the labeled $\Delta E_Q$-Fe(III) (blue) and $\Delta E_Q$-Fe(II) (maroon). The black open circles represent the measured data, while the black solid lines indicate the fitted data, which is composed of two pairs of doublets: one from the Fe(II) doublet (solid maroon lines) and another from the Fe(III) doublet (solid blue lines). **b**, The $^{57}$Fe Mössbauer spectra of 340nm_FP/Li$_{0.32}$FP/Li$_{1.0}$FP with the labeled $\Delta E_Q$-Fe(III) (blue) and $\Delta E_Q$-Fe(II) (maroon). **c**, Comparisons of $\Delta E_Q$-Fe(III) (blue) and $\Delta E_Q$-Fe(II) (maroon) between 20nm_Li$_X$FP and 340nm_Li$_X$FP particles. The sparse line patterns indicate the $\Delta E_Q$ of only one pair of meaning doublet is summarized for the end phases FP or Li$_{1.0}$FP.